# Metric Signature Transitions in Optical Metamaterials

Igor I. Smolyaninov [1] and Evgenii E. Narimanov [2]

[1] *Department of Electrical and Computer Engineering, University of Maryland, College Park, MD 20742, USA*

[2] *Birck Nanotechnology Center, School of Computer and Electrical Engineering, Purdue University, West Lafayette, IN 47907, USA*

**We demonstrate that the extraordinary waves in indefinite metamaterials experience ( - - + + ) effective metric signature. During a metric signature change transition in such a metamaterial, a Minkowski space-time is "created" together with large number of particles populating this space-time. Such metamaterial models provide a table top realization of metric signature change events suggested to occur in Bose-Einstein condensates and quantum gravity theories.**

PACS numbers: 78.20.Ci, 42.25.Bs, 71.36.+c, 78.30.-j.

The unprecedented degree of control of the local dielectric permittivity $\varepsilon_{ik}$ and magnetic permeability $\mu_{ik}$ tensors in electromagnetic metamaterials has fueled recent explosion in novel device ideas, and resulted in discovery of new physical phenomena. Advances in experimental design and theoretical understanding of metamaterials greatly benefited from the field theoretical ideas developed to describe physics in curvilinear space-time. Electromagnetic cloaking [1,2,3] and electromagnetic wormholes [4] may be cited as good examples. With the new found freedom of $\varepsilon_{ik}$ and $\mu_{ik}$ manipulation, experimentalists may create optical models of virtually any metric allowed in general relativity [5]. Recent literature [6-9] provides good examples of how it is being achieved in the case of optical analogues of black holes. On the other hand, compared to standard general relativity, metamaterial optics gives more freedom to design an effective space-time with very unusual properties. Light propagation in all static general relativity situations can be mimicked with positive $\varepsilon_{ik} = \mu_{ik}$ [10], while the allowed parameter space of the metamaterial optics is broader. Thus, the flat Minkowski space-time with the usual (-,+,+,+) signature does not need to be a starting point. Other effective signatures, such as the "two times" (2T) physics (-,-,+,+) signature may be realized [11]. Theoretical investigation of the 2T higher dimensional space-time models had been pioneered by Paul Dirac [12]. More recent examples can be found in [13,14]. Metric signature change events (in which a phase transition occurs between say (-,+,+,+) and (-,-,+,+) space-time signature) are being studied in Bose-Einstein condensates and in some modified gravitation theories (see ref.[15], and the references therein). In general, it is predicted that a quantum field theory residing on a spacetime undergoing a signature change reacts violently to the imposition of the signature change. Both the total number and the total energy of the particles generated in a signature change event are formally infinite [15]. Therefore, such a metric signature transition can be called a "Big Flash", which shares some similarities with the cosmological "Big Bang". A metamaterial model of a metric signature change event should be extremely interesting to observe. Unlike usual phase transitions, in which the physical system changes while the background metric is intact, the signature change transition affects the underlying background metric experienced by the system. Therefore, signature change events constitute a new kind of phase transition.

The equation of motion of a quantum scalar field $\varphi$ residing on some space-time is described by the Klein-Gordon equation [16]:

$$\hat{P}^i \hat{P}_i \varphi = m^2 \varphi \qquad (1)$$

where $\hat{P}^i = \partial/\partial x_i$ is the momentum operator. For a massless field in a "flat" (2+2) four-dimensional 2T space-time the Klein-Gordon equation can be written as

$$\left(\frac{\partial^2}{\partial x_1^2} + \frac{\partial^2}{\partial x_2^2} - \frac{\partial^2}{\partial x_3^2} - \frac{\partial^2}{\partial x_4^2}\right)\varphi = 0 \qquad (2)$$

in the coordinate space, and as

$$\left(-k_1^2 - k_2^2 + k_3^2 + k_4^2\right)\varphi_k = 0 \qquad (3)$$

in the $k$-space. To illustrate how dynamics described by eq.(2) can be mimicked in metamaterials, let us start with a non-dispersive (if the operating bandwidth is sufficiently narrow the dispersion can be neglected) and non-magnetic uniaxial anisotropic material with dielectric permittivities $\varepsilon_x = \varepsilon_y = \varepsilon_1$ and $\varepsilon_z = \varepsilon_2$. The wave equation in such a material can be written ([17]) as

$$-\frac{\partial^2 \vec{E}}{c^2 \partial t^2} = \vec{\varepsilon}^{-1} \vec{\nabla} \times \vec{\nabla} \times \vec{E} \qquad (4)$$

where $\vec{\varepsilon}^{-1}$ is the inverse dielectric permittivity tensor calculated at the center frequency of the signal bandwidth. Any electromagnetic field propagating in this material can be expressed as a sum of the "ordinary" and "extraordinary" contributions, each of these being a sum of an arbitrary number of plane waves polarized in the "ordinary" ($\vec{E}$ perpendicular to the optical axis) and "extraordinary" ($\vec{E}$ parallel to the plane defined by the k–vector of the wave and the optical axis) directions. Let us defined our "scalar" extraordinary wave function as $\varphi = E_z$ (so that the ordinary portion of the electromagnetic field does not contribute to $\varphi$). Equation (4) then yields the following form of the wave equation:

$$\frac{\partial^2 \varphi}{c^2 \partial t^2} = \frac{\partial^2 \varphi}{\varepsilon_1 \partial z^2} + \frac{1}{\varepsilon_2}\left(\frac{\partial^2 \varphi}{\partial x^2} + \frac{\partial^2 \varphi}{\partial y^2}\right) \quad (5)$$

While in ordinary crystalline anisotropic media both $\varepsilon_1$ and $\varepsilon_2$ are positive, this is not necessarily the case in metamaterials. In the indefinite metamaterials considered for example in [18,19] $\varepsilon_1$ and $\varepsilon_2$ have opposite signs. These metamaterials are typically composed of multilayer metal-dielectric or metal wire array structures, as shown in Fig.1(a). Optical properties of such metamaterials are quite unusual. For example, there is no usual diffraction limit in an indefinite metamaterial [18-23]. In the absence of dispersion, eq.(5) in the case of $\varepsilon_1 < 0$ and $\varepsilon_2 > 0$ looks like the Klein-Gordon equation (2) for a massless field in a "flat" (2+2) four-dimensional 2T space-time.

However, metamaterials generally show high dispersion, which often cannot be neglected. In this case eq.(4) is replaced by:

$$\frac{\omega^2}{c^2}\vec{D}_\omega = \vec{\nabla} \times \vec{\nabla} \times \vec{E}_\omega \text{ and } \vec{D}_\omega = \vec{\varepsilon}_\omega \vec{E}_\omega \quad (6)$$

which results in the following wave equation for $\varphi_\omega$:

$$-\frac{\omega^2}{c^2}\varphi_\omega = \frac{\partial^2 \varphi_\omega}{\varepsilon_1 \partial z^2} + \frac{1}{\varepsilon_2}\left(\frac{\partial^2 \varphi_\omega}{\partial x^2} + \frac{\partial^2 \varphi_\omega}{\partial y^2}\right) \quad (7)$$

or equivalently in the k-space:

$$\frac{\omega^2}{c^2}\varphi_{\omega,k} = \frac{k_z^2}{\varepsilon_1}\varphi_{\omega,k} + \frac{k_x^2 + k_y^2}{\varepsilon_2}\varphi_{\omega,k} \quad (8)$$

Let us consider the case of $\varepsilon_1 < 0$ and $\varepsilon_2 > 0$ and assume that this behavior holds in some frequency range. Here we observe that the effective space-time signature as seen by extraordinary light propagating inside the metamaterial has different character depending on the frequency. At high frequencies (above the plasma frequency of the metal), the metamaterial exhibits "normal" Minkowski effective metric with (-,+,+,+) signature, while at low frequencies the metric signature changes to (-,-,+,+). After simple coordinate transformation eq.(8) can be re-written as

$$\left(-k_t^2 - k_z^2 + k_x^2 + k_y^2\right)\varphi_{\omega,k} = 0 \qquad (9)$$

which coincides with the Klein-Gordon equation for a (2+2) space-time in the k-space (see eq.(3)). Alternatively, in the case of $\varepsilon_1 > 0$ and $\varepsilon_2 < 0$, eq.(8) can be re-written as

$$\left(k_t^2 + k_x^2 + k_y^2 - k_z^2\right)\varphi_{\omega,k} = 0 \qquad (10)$$

As a result, both at the small and the large frequencies the effective metric looks like the Minkowski space-time. However, at small frequencies the *z*-coordinate assumes the role of a time-like variable. Note that causality and the form of eqs.(9) and (10) place stringent limits on the material losses and dispersion of hyperbolic metamaterials: a dispersionless and lossless hyperbolic metamaterial would violate causality. On the other hand, such metamaterials indeed enable experimental exploration of the metric signature transitions [15] to and from the Minkowski space-time. Unlike the infinite number of particles created in an idealized theoretical signature change event [15], metamaterial losses limit the number of particles created during the real signature change transition. Nevertheless, our calculations indicate that the number of created particles is still very large (see Fig.3).

A composite metamaterial, which exhibits a metric signature phase transition as a function of temperature can be designed using materials which exhibit metal-dielectric transitions. A numerical example of such a metamaterial is presented in Fig.1. It is making use of a pronounced change in dielectric constant of gallium upon phase transition from the crystalline semi-metallic α-gallium phase to the liquid phase, in which gallium behaves as a low-loss free-electron metal. The melting point of gallium at the atmospheric pressure is located at 29.8° C [24], so the "metric signature phase transition" in such a metamaterial is easy to observe. This transition may also be induced by external femtosecond light pulses, as has been observed in [25] for gallium films on silica substrates. The wired metamaterial structure in the simulation is assumed to be made of thin gallium wires distributed inside the silica matrix, so that observations of ref.[25] can be used. The real and imaginary parts of the dielectric constant of liquid gallium were taken from ref.[24]. The optical constants of solid α-gallium were taken from [26]. The Maxwell-Garnett approximation has been used to calculate the real and imaginary parts of $\varepsilon_z$ and $\varepsilon_{x,y}$ of the "wired" metamaterial for the liquid and solid states of the gallium wires (Fig.1(b,c)). It has been validated in [19] in the case of low volume fraction $N$ of the metallic component of the metamaterial. Therefore, the case of $N_{Ga}$=0.1 volume fraction of gallium has been considered. It is possible to further refine the Maxwell-Garnet approximation by taking into account dynamical interaction between the wires [27]. Our simulations indicate that the extraordinary photons propagating inside the wired Ga-based metamaterial experience the metric signature change from the Minkowski-like (-,+,+,+) to the 2T physics (-,-,+,+) upon melting of the Ga wires. Moreover, the losses in the (-,-,+,+) regime remain relatively low: at 1.8 eV the ratio of real to imaginary part of $\varepsilon_z$ is about $Re(\varepsilon_z)/Im(\varepsilon_z)$~2, which is translated into about factor of 4 ratio between the real and imaginary parts of the photon wavevector. In principle, propagation loss may be further improved by using a gain medium such as the dye-doped silica instead of passive silica glass [28]. We also note that the "layered" Ga-based metamaterial having $N_{Ga}$=0.1 would also demonstrate a metric phase transition as a function of temperature. Femtosecond light-induced phase transition inside such a metamaterial (which according to ref.[25] occur extremely fast) would look like a version of the dynamical Casimir effect experiment [29,30], which is predicted to produce a "flash" of Casimir light emission.

The "Big Flash" behavior is not limited to the artificial metal-dielectric metamaterials only. Many dielectric crystals, such as α-quartz, have lattice vibration modes which carry an electric dipole moment. The dipole moment couples the lattice vibrations to the radiation field in the crystal to form the phonon-polariton modes [31]. As a result, multiple reststrahlen bands are formed near the frequencies of the dipole-active lattice vibration modes $\omega_n$ (where $n$ is the mode number) in which both $\varepsilon_1$ and $\varepsilon_2$ become metal-like and negative. These bands are typically located in the mid-IR region of the electromagnetic spectrum. Due to crystal anisotropy, $\varepsilon_1$ and $\varepsilon_2$ change sign at slightly different frequencies [32]. Thus, in the narrow frequency ranges near the boundaries of the reststrahlen bands $\varepsilon_1$ and $\varepsilon_2$ have different signs. In these frequency ranges natural crystals behave as indefinite metamaterials, and the extraordinary light dispersion law is either

$$k_x^2 + k_y^2 - k_z^2 = \omega_n^2/c^2 \quad , \text{ or} \qquad (11)$$

$$k_z^2 - k_x^2 - k_y^2 = \omega_n^2/c^2 \qquad (12)$$

The latter case corresponds formally to the free particle spectra in a (2+1) dimensional Minkowski space-time, in which the mass spectrum is given by the spectrum $\omega_n$ of the lattice vibrations (Fig.2). It is interesting to note that the liquid-solid phase transition in such a crystal provides us with an example of a phase transition in which a formal (2+1) dimensional Minkowski space-time emerges together with a discrete free particle spectrum. The characteristic feature of this phase transition appears to be a "Big Flash" due to the sudden emergence of the infinities of the photonic density of states near the $\omega_n$ frequencies. Unlike the usual finite black body photonic density of states

$$\frac{dn}{d\omega} = \frac{\omega^2}{2\pi^2 c^3} \qquad (13)$$

defined by the usual $k_x^2 + k_y^2 + k_z^2 = \omega_n^2/c^2$ photon dispersion law in vacuum, the density of states of the extraordinary photons near the $\omega_n$ frequencies diverges in the lossless continuous hyperbolic medium limit:

$$\frac{dn}{d\omega} \approx \frac{K^3_{max}}{12\pi^2} \left| \frac{\varepsilon_2}{\varepsilon_1} \left( \frac{1}{\varepsilon_1} \frac{d\varepsilon_1}{d\omega} - \frac{1}{\varepsilon_2} \frac{d\varepsilon_2}{d\omega} \right) \right| \qquad (14)$$

where $K_{max}$ is the momentum cutoff [33]. $K_{max}$ is defined by either metamaterial structure scale, or by losses. As a result of the suddenly emerging divergences in the photonic density of states near the $\omega_n$ frequencies, these states are quickly populated during the liquid-solid phase transition. This event can be considered as a simultaneous emergence of the (2+1) dimensional Minkowski space-time and a very large number of particles (extraordinary photons), which quickly populate the divergent density of states, and hence the emergent (2+1) Minkowski space-time itself. The number of photons created as a result of such event is defined by the Maxwell–Boltzmann distribution at the temperature $T$ of the liquid-solid transition: $N \sim (dn/d\omega)\exp(-\hbar\omega_n/kT) \sim K^3_{max}\exp(-\hbar\omega_n/kT)$. Since for the reststrahlen bands $\hbar\omega_n \leq kT$, a large number of photons is created. Intriguingly, flashes of light are indeed observed during fast crystallization of many dielectric materials, which exhibit phonon-polariton resonances. This phenomenon is known since 18[th] century as crystalloluminescence [34]. Unfortunately, these observations may only be treated as the "Big Flash" events if similar flashes are observed in the mid IR range, or some mechanism of photon up-conversion will be confirmed.

On the other hand, in the case of $\hbar\omega_n \gg kT$ where $\omega_n$ is the frequency within the hyperbolic region of the dispersion law, the number of photons emitted during the metric signature change transition can be calculated via the dynamical Casimir effect. The total energy $E$ of emitted photons from a phase changing volume $V$ depends on the photon dispersion laws $\omega(k)$ in both media (see eq.(3) from ref.[30]):

$$\frac{E}{V} = \int \frac{d^3\vec{k}}{(2\pi)^3}\left(\frac{1}{2}\hbar\omega_1(k) - \frac{1}{2}\hbar\omega_2(k)\right) \qquad (15)$$

where $\omega_1(k)$ and $\omega_2(k)$ are the photon dispersion relations inside the respective media. Equation (15) is valid in the "sudden change" approximation, in which the dispersion law is assumed to change instantaneously. The detailed discussion of the validity of this approximation can be found in ref. [30]. Therefore, the number of photons per frequency interval emitted during the transition can be written as

$$\frac{dN}{Vd\omega} = \frac{1}{2}\left(\frac{dn_1}{d\omega} - \frac{dn_2}{d\omega}\right) \qquad (16)$$

where $dn_i/d\omega$ are the photonic densities of states inside the respective media. Since the photonic density of states diverges in the hyperbolic regions of the photon dispersion near the frequencies $\omega_n$ of the phonon-polariton resonances, the number of photons emitted during the transition would also diverge near $\omega_n$ in the lossless continuous medium limit, leading to the "Big Flash" behavior. Using eqs.(15,16) we can calculate the number of photons emitted during the metric signature phase transition in the gallium-based metamaterial presented in Fig.1, assuming that it is induced by a femtosecond light pulse [25], so that the "sudden change" approximation is valid.

Results of our numerical calculations of the photon energy d$E/d\omega$ (normalized per unit volume) emitted during the light-induced metric signature phase transition in the gallium-based metamaterial from Fig.1 are presented in Fig.3. The metamaterial structure in these calculations is assumed to be fine enough, so that the value of $K_{max}$ is defined by metamaterial losses. Thus, a dedicated experiment performed on an artificial or natural hyperbolic metamaterial exhibiting a metric signature phase transition is possible, and would be an extremely interesting development. In addition to Ga-based metamaterial geometry described above, such materials as vanadium oxide or other conducting transparent oxides materials (e.g. ITO, AZO and similar [35,36]) can be used as the phase-changing component of the hyperbolic metamaterial. These materials can be switched from dielectric to metal phase by injecting more free charge careers, and thus changing the plasma frequency either optically or electrically.

**Figure Captions**

**Figure 1.** (a) Schematic views of the "wired" and "layered" metamaterial structures. (b) and (c) show calculated real and imaginary parts of $\varepsilon$ of the composite wired metamaterial made of silica glass and gallium wires (Im$\varepsilon_{x,y}$<<Im$\varepsilon_z$, not shown). This medium demonstrates a metric phase transition in the visible frequency range when gallium wires are converted from the solid to liquid state. The case of $N_{Ga}$=0.1 volume fraction of gallium has been considered in these simulations. The range of frequencies for which the effective metric signature changes from the Minkowski (-,+,+,+) to the 2T physics (-,-,+,+) is shown by the dashed lines.

**Figure 2.** Hyperbolic dispersion relation allowing unbounded values of the wavevector (blue arrow) due to which the photonic density of states diverges. The group velocity vectors (red arrow) lie within a cone which implies light propagation in such media is inherently directional.

**Figure 3.** Photon energy $dE/d\omega$ (normalized per unit volume) emitted during the femtosecond light-induced metric signature phase transition in the gallium-based metamaterial presented in Fig.1.

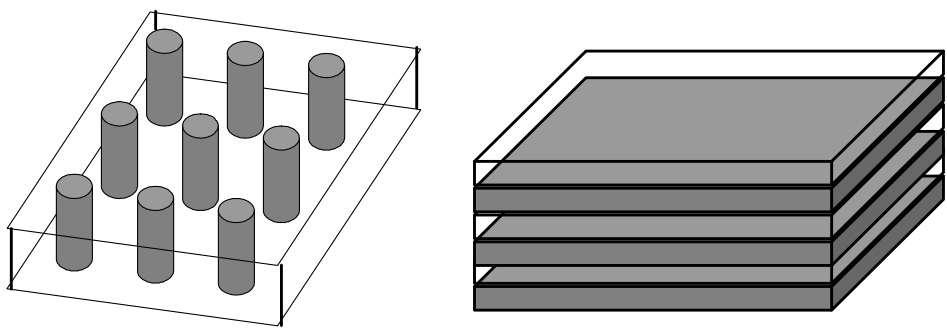

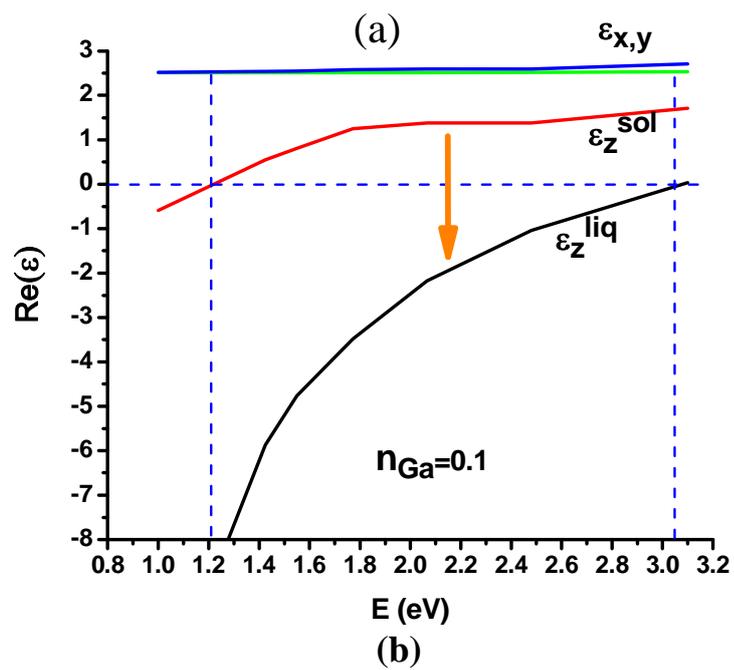

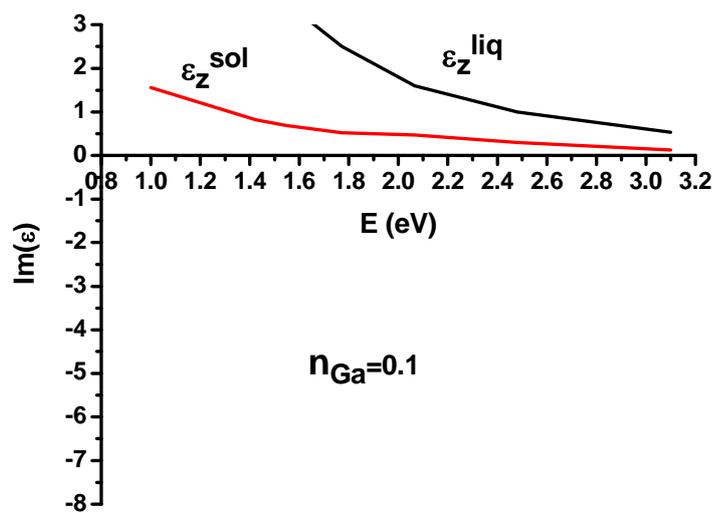

Fig. 1

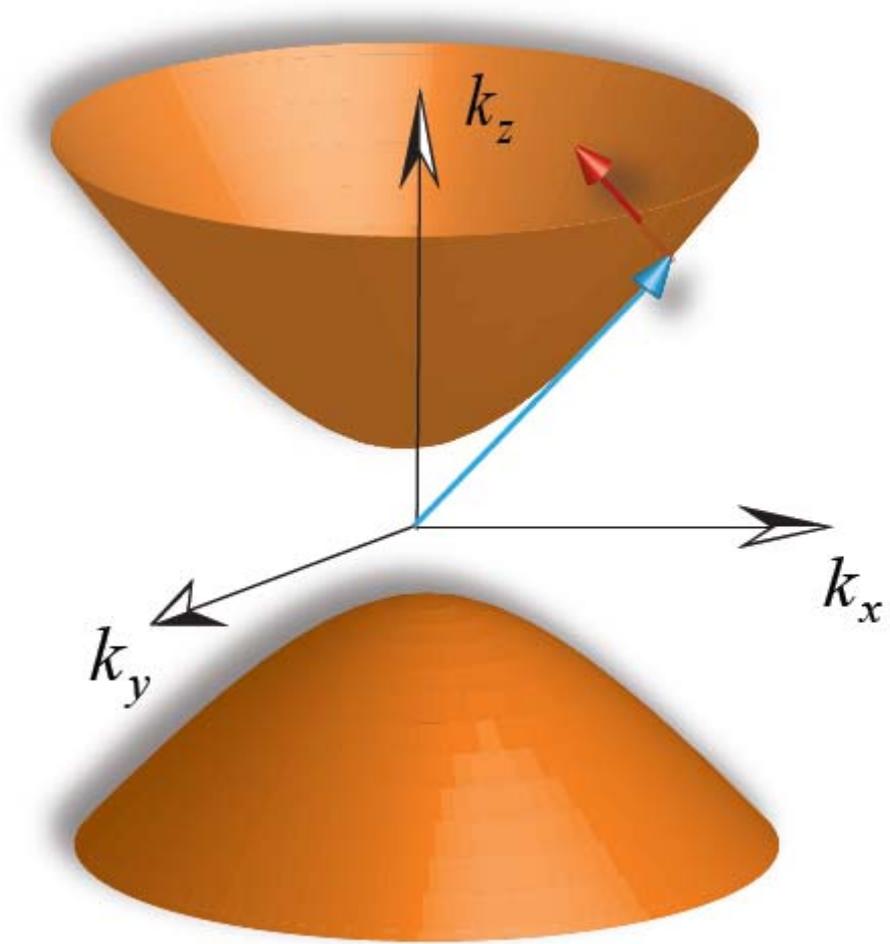

Fig.2

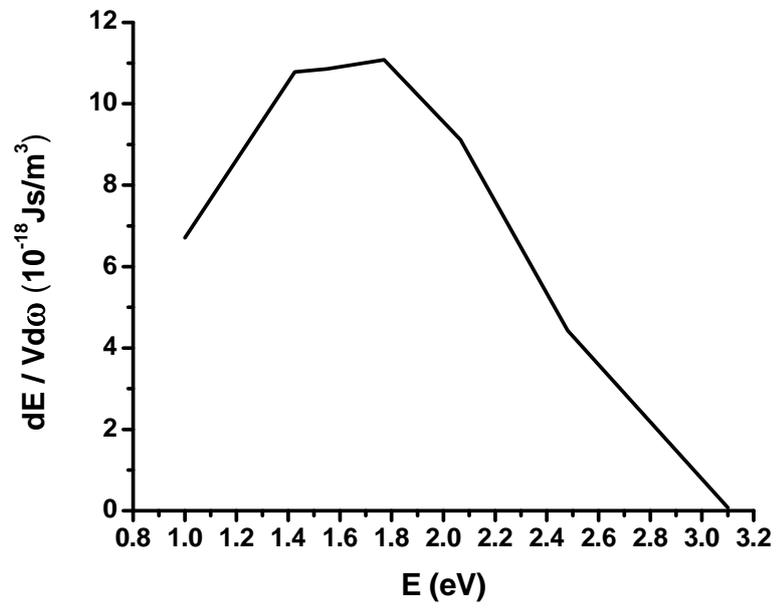

Fig.3